\documentclass[12pt]{article} 

\usepackage{graphics,color}
\usepackage{amssymb}
\usepackage{amsbsy}
\usepackage{epsfig}

\newcommand{\be}{\begin{equation}}
\newcommand{\ee}{\end{equation}}
\newcommand{\bea}{\begin{eqnarray}}
\newcommand{\eea}{\end{eqnarray}}
\newcommand{\bit}{\begin{itemize}}
\newcommand{\eit}{\end{itemize}}
\newcommand{\bra}{\langle}
\newcommand{\ket}{\rangle}
\newcommand{\zv}{{\mathbf 0}}
\newcommand{\si}{\mbox{$\sum$}\hs{-0.45cm}\int}
\newcommand{\sgn}{\mbox{sgn}}
\newcommand{\im}{{\mathrm{Im}}}
\newcommand{\re}{{\mathrm{Re}}}

\newcommand{\tr}{{\mathrm{tr}}}
\newcommand{\cO}{{\cal O}}
\newcommand{\cP}{{\cal P}}
\newcommand{\cD}{{\cal D}}

\newcommand{\om}{\omega}
\newcommand{\omp}{\omega_{+}}
\newcommand{\omm}{\omega_{-}}
\newcommand{\ompm}{\omega_{\pm}}
\newcommand{\tw}{\Gamma}
\newcommand{\aste}{{}^\ast\!}

\newcommand{\po}{p^{0}}
\newcommand{\qo}{q^{0}}
\newcommand{\ko}{k^{0}}
\newcommand{\di}{\int}
\newcommand{\gv}{\boldsymbol\gamma}
\newcommand{\gz}{\gamma^{0}}
\newcommand{\gi}{\gamma^{i}}
\newcommand{\pv}{{\mathbf p}}
\newcommand{\kv}{{\mathbf k}}
\newcommand{\qv}{{\mathbf q}}
\newcommand{\rv}{{\mathbf r}}
\newcommand{\vv}{{\mathbf v}}
\newcommand{\pvuni}{\hat{{\mathbf p}}}
\newcommand{\kvuni}{\hat{{\mathbf k}}}

\newcommand{\rvuni}{\hat{{\mathbf r}}}
\newcommand{\vvuni}{\hat{{\mathbf v}}}
\newcommand{\puni}{\hat{p}}
\newcommand{\kuni}{\hat{k}}

\newcommand{\lmax}{\Lambda_{max}}
\newcommand{\lmin}{\Lambda_{min}}
\newcommand{\lmg}{\ln\!\left(\frac{m_{D}}{\lmin}\right)}

\newcommand{\hm}{\hspace*{-0.6cm}}
\newcommand{\hs}[1]{\hspace*{#1}}
\newcommand{\hp}{\hspace*{1cm}}

\newcommand{\al}{\alpha}

\newcommand{\bean}{\begin{eqnarray*}}
\newcommand{\eean}{\end{eqnarray*}}
\newcommand{\nn}{\nonumber}
\newcommand{\veck}{{\mathbf k}}
\newcommand{\vecp}{{\mathbf p}}

\newcommand{\vecr}{{\mathbf r}}

\newcommand{\vecnul}{{\mathbf 0}}
\newcommand{\iop}{i0^{+}}

\begin{document}

\title{
\vskip -100pt
{
\begin{normalsize}
\mbox{} \hfill  hep-ph/0209048
\vskip  70pt
\end{normalsize}
}
Ward identity and electrical conductivity in hot QED
}
\author{
Gert Aarts\thanks{
email: aarts@mps.ohio-state.edu}\addtocounter{footnote}{1}
{} and
J.\ M.\ Mart{\'\i}nez Resco\thanks{
email: marej@mps.ohio-state.edu}\addtocounter{footnote}{2}\\
{}\\
\normalsize
{\em Department of Physics, The Ohio State University}\\
\normalsize
{\em 174 West 18th Avenue, Columbus, OH 43210, USA}
}
\date{September 4, 2002}
\maketitle

\begin{abstract}
We study the Ward identity for the effective photon-electron vertex
summing the ladder diagrams contributing to the electrical conductivity in
hot QED at leading logarithmic order. It is shown that the Ward identity
requires the inclusion of a new diagram in the integral equation for the
vertex that has not been considered before. The real part of this 
diagram is subleading and therefore the final expressions for the
electrical conductivity at leading logarithmic order are not affected.
\end{abstract}

\newpage


\section{Introduction}

Transport coefficients in quantum field theories at finite temperature
have received an increasing amount of attention over the last few years,
not only because of their potential relevance in some physical
environments, such as heavy-ion collisions and the early universe, but
also because, from a theoretical point of view, their calculation turns
out to be highly nontrivial. A perturbative analysis can be used when the
temperature is sufficiently high and the theory is weakly coupled.  
However, the computation of transport coefficients in hot gauge theories
within the framework of thermal field theory remains a difficult task due
to the necessity of summing an infinite number of Feynman diagrams,
so-called ladder diagrams~\cite{smilga}. This has favoured the use of
effective descriptions such as transport theory
\cite{baym,heiselberg,amy-tc}. Another alternative is the use of lattice
field theory \cite{Karsch:1986cq}, which allows one in principle to obtain
transport coefficients at temperatures where a perturbative analysis
(either with field or transport theory) is not valid.  This approach has
not been completely developed and presents its own
difficulties~\cite{aarts}.

It is within the kinetic approach that it was first realized that
screening processes in the plasma at the scale of the Debye mass are
enough to render results finite~\cite{baym}. The first complete
calculation of transport coefficients in hot gauge theories at leading
logarithmic order appeared only recently~\cite{amy-tc}, also using kinetic
theory. For a scalar theory the ladder diagrams have been summed
explicitly by Jeon~\cite{jeon-1} using a Bethe-Salpeter equation for an
effective vertex and the leading-order results for the shear and bulk
viscosities have been obtained. The conclusions of his diagrammatic
analysis have been confirmed in Refs.\ \cite{Wang:1999gv}. Furthermore,
Jeon and Yaffe~\cite{jeon-2} showed the equivalence between the
diagrammatic and the kinetic approach: to leading order the linearized
Boltzmann equation for the distribution function and the Bethe-Salpeter
equation for the effective vertex yield equivalent results. For QCD, a
simplified ladder summation \cite{yo} reproduces the result for the color
conductivity at leading logarithmic order~\cite{Bodeker:1998hm}.

\begin{figure}[t] 
 \centerline{\epsfig{figure=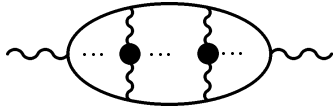,width=5cm}}
 \caption{Typical ladder diagram contributing to the electrical
 conductivity. The side rails are hard, nearly on-shell fermions and the
 rungs are soft photons.}
 \label{figladders}
\end{figure}

\begin{figure}[h]
 \centerline{\epsfig{figure=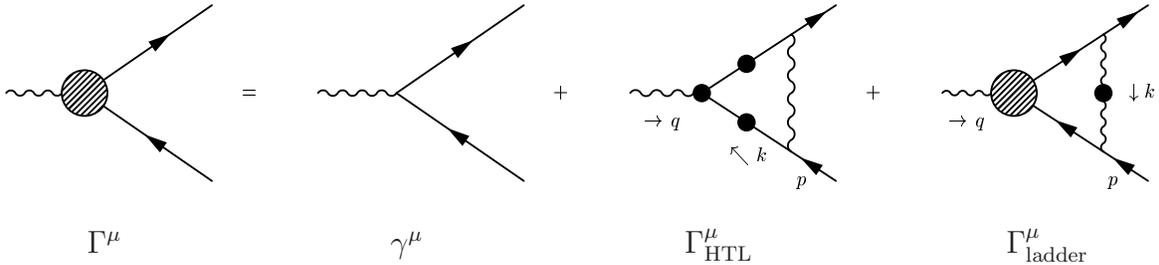,width=16cm}}
 \centerline{ 
 $\Gamma^\mu$ \hspace{3.3cm} $\gamma^\mu$ 
 \hspace{3.2cm} $\Gamma_{\rm HTL}^\mu$  
 \hspace{3.1cm} $\Gamma_{\rm ladder}^\mu$ }
 \caption{Integral equation for the effective photon-electron vertex 
 function $\Gamma^\mu$. The second diagram on the right-hand-side with a 
 hard photon and HTL vertex and fermion propagators is new and is required 
 to fulfill the Ward identity.}
 \label{figie}
\end{figure}

Only very recently a simple and economical way of summing the ladder
series via a Bethe-Salpeter equation in the imaginary-time formalism
has been presented by Valle Basagoiti~\cite{valle}, for both scalar and
(non)abelian gauge theories. To leading logarithmic order, the integral
equations obtained in Ref.~\cite{valle} are identical to those found
previously in the kinetic approach~\cite{amy-tc}. However, for gauge
theories the integral equations for the effective vertices used in
Ref.~\cite{valle} are not consistent with the Ward identities. In the case
of the electrical conductivity in QED, which we will consider in this
paper, this can be understood as follows. As usual, the photon-electron
vertex and the fermion propagator are related via the Ward identity. A
typical ladder diagram contributing to the electrical conductivity at
leading logarithmic order is shown in Fig.~\ref{figladders}.  Propagators
for the nearly on-shell fermions on the side rails with hard momentum ($p
\equiv |\pv| \sim T$, with $T$ the temperature) have to include the
fermionic thermal width, such that singularities due to so-called pinching
poles are regulated. This thermal width receives contributions from
processes involving both a soft ($p\sim eT$) photon and a soft fermion.
Ladder diagrams as the one shown in Fig.~\ref{figladders} can be summed by
introducing an effective photon-electron vertex involving a soft photon
rung~\cite{valle}.
One expects that the Ward identity relates the contribution to the thermal
width from soft photons to the vertex with a soft photon rung. However,
the contribution to the thermal width from soft fermions, appearing at
order $e^{4}T\ln(1/e)$, has no counterpart in the equation for the vertex
function presented in Ref.~\cite{valle}. Therefore, the Ward identity is
not fulfilled and the equation for the effective vertex given in
Ref.~\cite{valle} cannot be complete. We show in this paper that in order
to satisfy the Ward identity a new diagram involving soft fermions has to
be included, so that the integral equation is the one depicted in Fig.\ 
\ref{figie}. As far as we know, this diagram has not been discussed 
before.

Concerning the electrical
conductivity, however, only the real part of the effective photon-electron
vertex is required. It turns out that the real part of the new diagram is
parametrically suppressed with respect to the tree-level vertex. Therefore
we find that the presence of the vertex correction involving soft fermions
does not affect the final result for the electrical conductivity at
leading logarithmic order.

The paper is organized as follows. In Sec.~\ref{ec} we review the
derivation of the electrical conductivity in terms of a particular
analytic continuation of the effective vertex of Ref.~\cite{valle}. The
complete thermal width of order $e^{4}T\ln(1/e)$ for an on-shell electron
with hard momentum is computed in Sec.~\ref{secwidth}. In Sec.~\ref{ward}
we show the consistency of the modified vertex equation with the Ward
identity. In Sec.~\ref{de} we show that the new integral equation leads to
the same leading-log differential equation as in Refs.~\cite{amy-tc,valle}
for that piece of the effective vertex relevant for the electrical
conductivity.  Conclusions are presented in Sec.~\ref{seccon}. We have
summarized convenient sum rules in Appendix~\ref{appsumrules}. The
calculation of the new diagram is detailed in Appendix~\ref{spatial}.


\section{Electrical conductivity}   
\label{ec}

The Kubo formula for the electrical conductivity in QED is 
\be
 \label{eqkubo}
 \sigma = \frac{1}{6}\, 
 \frac{\partial}{\partial \qo}\rho(\qo,\zv)\Big|_{\qo=0},
\ee
where $\rho$ is the spectral density associated with the spatial part 
of the retarded polarization tensor
\be
 \rho(q^{0},\qv) = 2\im\,\Pi^{ii}_{R}(q^{0},\qv), \hp  
 \Pi^{ii}_{R}(x-y) = i\theta(x^{0}-y^{0})\bra[j^{i}(x),j^{i}(y)]\ket,
\ee
with $j^i(x) = \bar\psi(x)\gamma^i\psi(x)$ the electromagnetic current.  
The retarded correlator can be obtained from the Euclidean one by
analytical continuation,
\be
 \Pi^{ii}_{R}(q^{0},\qv) = \Pi^{ii}_E(i\om_{q}\to\qo+\iop,\qv),
\ee
with $\om_{q}=2\pi nT$ ($n\in \mathbb{Z}$) the Matsubara frequency. 
The relevance of ladder diagrams for the conductivity can be understood as
follows. We start with the simple one-loop expression: since in the
Kubo formula (\ref{eqkubo}) the correlator appears with vanishing external
momentum, the fermionic propagators in the one-loop expression share
almost the same momentum and so-called pinching poles are present. They
cause the one-loop contribution to diverge unless the thermal width is
present in the electron propagators~\cite{smilga}. Because the dominant
contribution arises when the electrons are on-shell and carry hard
momentum, the width is included by replacing the Dirac delta functions of
the free single-particle spectral densities with Lorentzian spectral
functions\footnote{We assume the temperature and hence the hard fermion
momentum is sufficiently high such that both the zero-temperature electron
mass and the real part of the fermionic self-energy can be safely 
neglected.}
\be
 \rho^{\rm{free}}_{\pm}(\om,\pv) = 2\pi\delta(\om \mp p)\longrightarrow
 \rho_{\pm}(\om,\pv) = \frac{\tw_{\pv}}{(\om\mp p)^{2}+(\tw_{\pv}/2)^{2}},
\ee
where $\tw_{\pv}$ is the thermal width of a fermion with hard on-shell
momentum. These positive- and negative-energy spectral densities are
related to the electron propagator as
\be
 S(\po,\pv) = 
 \Delta_{+}(\po,\pv)h_{+}(\pvuni)+\Delta_{-}(\po,\pv)h_{-}(\pvuni),
 \;\;\;\;\;\;\;
 \Delta_{\pm}(\po,\pv) = 
 -\int\frac{d\om}{2\pi}\frac{\rho_{\pm}(\om,\pv)}{\po-\om},
\ee
with\footnote{The gamma-matrices obey 
$\{\gamma^{\mu},\gamma^{\nu}\} =
2g^{\mu\nu}$ with $g^{\mu\nu} = \mbox{diag}(1,-1,-1,-1)$.}
\be \label{eqhp}
 h_{+}(\pvuni) = \frac{1}{2}\left(\gz-\gv\cdot\pvuni\right) 
 = \sum_{\lambda}u_{\lambda}(\pvuni)\bar{u}_{\lambda}(\pvuni),
 \;\;\;\;\;\;\;
 h_{-}(\pvuni) = \frac{1}{2}\left(\gz+\gv\cdot\pvuni\right) 
 = \sum_{\lambda}v_{\lambda}(\pvuni)\bar{v}_{\lambda}(\pvuni), 
\ee
where $u_{\lambda}$ ($v_{\lambda}$) are spinors for the electron
(positron) in a simultaneous chirality-helicity base ($\lambda=\pm$ 
indicates the helicity, $\pvuni=\pv/p$). 
Similarly we write the self-energy as
\be
 \Sigma(\po,\pv) = 
\Sigma_{-}(\po,\pv)h_{+}(\pvuni)+\Sigma_{+}(\po,\pv)h_{-}(\pvuni).
\ee
The use of Lorentzian spectral densities leads to fermionic propagators 
\be \label{self}
  \Delta_{\pm}(z,\pv)=\frac{-1}{z\mp p-\Sigma_{\pm}(z,\pv)}, \hp 
  \Sigma_{\pm}(z,\pv)=-i\sgn\!\left[\im(z)\right]\frac{\tw_{\pv}}{2}.
\ee
This propagator has a cut on the real axis due to the discontinuity of the
sign function. In particular, the retarded and advanced propagators and 
self-energies for hard on-shell fermions are 
\bea
 \Delta^{R}_{\pm}(\po,\pv) = &&\hm \frac{-1}{\po\mp p+i\tw_{\pv}/2} 
 = \left[ \Delta^{A}_{\pm}(\po,\vecp)\right]^{*},\\
 \label{eqsigmaRA}
 \Sigma_{\pm}^{R}(\po, \pv) = &&\hm -i\tw_{\pv}/2 
 = \left[\Sigma_{\pm}^{A}(\po, \pv)\right]^{*},
\eea
when $\po\simeq \pm p$.
The presence of the width regulates the pinching-pole divergence in the
one-loop expression, which now behaves as $1/\tw_{\pv}$. However, the
immediate consequence is the need to sum all ladders diagrams with soft 
photon rungs, like
the one depicted in Fig.~\ref{figladders}. Since each new rung introduces
a pair of propagators with pinching poles and the width scales
(naively) as $e^2$, the powers of the coupling constant introduced by the
rung are compensated for by the factor $1/\tw_{\pv}$ from the
nearly-pinching poles. As a result it is necessary to sum all 
contributions from uncrossed ladders.\footnote{Actually, the 
thermal width or corresponding inverse time 
scale $\sim e^2T$ never appears in the calculation of the conductivity. 
Instead the relevant scale 
is $\Gamma_\pv \sim e^4T\ln(1/e)$. Therefore we think that a better way to 
justify the importance of ladder diagrams is as 
follows. For each additional soft photon rung, include a factor $e^2$ from 
the explicit interaction vertices, a factor $m_D^2\ln(1/e)$ from the 
integration over the rung, and a factor $\delta(\om\pm p)/\Gamma_\pv \sim
[e^4T^2\ln(1/e)]^{-1}$ from the additional pinching poles, see Eq.\ 
(\ref{eqpp}). Putting this together gives that the contribution of each 
additional rung is $\sim 1$ and all ladder diagrams are equally important.
}

These diagrams can be summed with a Bethe-Salpeter equation for an
effective vertex $\tw^{\mu}$.  In Ref.~\cite{valle} such an equation was
written and it was shown that the spatial part of the integral equation,
relevant for the transport coefficient, reduces to leading logarithmic
accuracy to a differential equation equivalent to the one obtained
previously in Ref.~\cite{amy-tc} using kinetic theory. 
As discussed in the Introduction, the equation for the vertex
presented in Ref.~\cite{valle} does not satisfy the Ward identity and can
therefore not be complete. In order for the Ward identity to be fulfilled
a new diagram has to be included such that the integral equation is the
one depicted in Fig.~\ref{figie}.

The Euclidean correlator summing all the ladder diagrams is then given
by\footnote{As we will see below it is sufficient to have one full 
($\Gamma^i$) and one bare ($\gamma^i$) vertex since the real 
part of $\Gamma_{\rm HTL}^{i}$ is subleading.}
\be
 \Pi_E^{ii}(Q)=e^{2}\si_{P} \tr\, \gamma^{i}S(P+Q)\Gamma^{i}(P+Q,P)S(P),
\ee
with $Q=(i\om_q,\vecnul)$.
We now follow Ref.~\cite{valle} to express the electrical conductivity in
terms of a particular analytic continuation of the effective vertex. After
doing the sum over Matsubara frequencies, only products of retarded and 
advanced fermion propagators $S^{R}(\po+\qo, \pv) S^{A}(\po, \pv)$ must be 
retained because only these can   
have pinching poles. Furthermore, since $\qo$ goes to zero, it cannot 
change the mass shell condition of the electrons on the side rails with hard 
momentum. Thus pinching poles arise only from the products 
$\Delta_{\pm}^{R}(\po+\qo,\pv)\Delta_{\pm}^{A}(\po,\pv)$ and we find
\bea  \label{eqsigma}
 \Pi_R^{ii}(\qo,\zv) = 2ie^{2}\int_{\pv,\om}
 \left[n_{F}(\om+\qo)-n_{F}(\om)\right] 
 \bigg[ \Delta^{R}_{+}(\om+\qo,\pv)\Delta^{A}_{+}(\om,\pv) 
 \puni^{i}D_{+}^{i}(\om+\qo,\om;\pv) && \nn  \\ 
 -\Delta^{R}_{-}(\om+\qo,\pv)\Delta^{A}_{-}(\om,\pv) 
 \puni^{i}D_{-}^{i}(\om+\qo,\om;\pv)\bigg],&&
\eea
where $n_F(\om) = 1/[\exp(\om/T)+1]$ is the Fermi distribution, and
\be
 \int_{\vecp,\om} = \int \frac{d^3p}{(2\pi)^3} \int \frac{d\om}{2\pi}.
\ee 
Here we used 
\be
h_{\pm}(\pvuni)\gi h_{\pm}(\pvuni) = \pm\puni^{i}h_{\pm}(\pvuni),
\ee
and defined
\bea
 \label{eqDplus}
 D^{\mu}_{+}(\om+\qo,\om ; \pv) \equiv &&\hm 
 \bar{u}_{\lambda}(\pvuni) \tw^{\mu}(\om+\qo+\iop,\om-\iop;\pv) 
 u_{\lambda}(\pvuni),  \\
 \label{eqDmin}
 D^{\mu}_{-}(\om+\qo,\om ; \pv) \equiv &&\hm 
 \bar{v}_{\lambda}(\pvuni) \tw^{\mu}(\om+\qo+\iop,\om-\iop;\pv)
 v_{\lambda}(\pvuni).
\eea
Both helicities give the same result such that the sum over helicities
yields a trivial factor 2 in Eq.~(\ref{eqsigma}).
Note that out of the many vertex functions with real energy
arguments \cite{Carrington:1996rx} only one particular
analytical continuation appears. Now, in the limit
$\qo\to 0$ and in the limit of narrow width (weak coupling), the pair 
of propagators goes to its pinching-pole limit,
\be
 \label{eqpp}
 \lim_{\qo\to 0} \Delta^{R}_{\pm}(\om+\qo,\pv)\Delta^{A}_{\pm}(\om,\pv) = 
 \frac{1}{(\om\mp p)^{2}+(\tw_{\pv}/2)^{2}}
 \longrightarrow \frac{2\pi}{\tw_{\pv}}\delta(\om\mp p),
\ee
forcing the on-shell condition $\om=\pm p$. Since in the pinching-pole
limit the product of propagators (\ref{eqpp}) is real and only the 
imaginary part of $\Pi_R^{ii}(\qo,\zv)$ is needed for the electrical 
conductivity, only the real part of the effective vertex contributes. 
Therefore we define 
\be
 \label{eqReD}
 \cD^{i}_{\pm}(\pv) \equiv \re\, D^{i}_{\pm}(\pm p+\qo,\pm p;\pv) 
\Big|_{\qo=0}.
\ee
Finally, since due to rotational invariance $\cD^{i}_{\pm}(\pv) = 
\puni^{i}\,\cD_{\pm}(p)$ and due to $CP$ invariance $\cD_{+}(p) = 
-\cD_{-}(p) \equiv \cD(p)$, the electrical conductivity is given by
\be
\label{eqsigmacon}
 \sigma=-\frac{4e^{2}}{3}\int_{\pv}n_{F}'(p)\frac{\cD(p)}{\tw_{\pv}}.
\ee
This expression can be easily compared with the result from kinetic
theory~\cite{amy-tc}. The factor 4 reflects that both electrons and 
positrons with either helicity contribute in the same way.


\section{Thermal Width}
\label{secwidth}

The electrical conductivity depends on the thermal width $\tw_{\pv}$ of a
hard on-shell fermion, which screens the pinching-pole singularity and
naturally sets an inverse time scale in the system. Kinetic theory
calculations \cite{baym,amy-tc} show that the relevant inverse relaxation
time for the electrical conductivity is $1/\tau \sim
e^{4}T\ln(T/m_{D})\sim e^{4}T\ln(1/e)$, coming from large angle scattering
between the hard nearly on-shell fermions in the plasma as well as from
scattering processes that change the type of excitation.  The thermal
width, on the other hand, is dominated by scattering processes in which
the fermions exchange a soft quasistatic transverse gauge boson (the
leading term is in fact logarithmically divergent, reflecting that in QED
the thermal width is ill-defined)  \cite{smilga,blaizot}. This dominant
contribution should therefore not be relevant for the calculation of the
electrical conductivity to leading logarithmic order. This is indeed what
is found in Refs.~\cite{smilga,valle} and will be confirmed in
Section~\ref{de}.\footnote{Note that in the case of the shear viscosity in
a scalar theory or color conductivity in QCD the scattering processes that
give the relevant relaxation time are those that also dominate the thermal
width. In these cases the simple relation $1/\tau_{\pv} \sim \tw_{\pv}$
holds.} The thermal width, however, receives subleading contributions from
scattering regimes different than the previous one. A contribution of
order $e^{4}T\ln(1/e)$ arises from the one-loop diagram with a soft
fermion (see the second diagram in Fig.~\ref{figselfenergy} below) and has
been computed in Ref.~\cite{valle}. This contribution corresponds to
Compton scattering and pair annihilation/creation processes, as can be
seen by cutting the diagram, which are mediated by a soft fermion screened
at the scale of the Debye mass. As is shown in this section, there is also
a contribution to the thermal width of order $e^4T\ln(1/e)$ from the
one-loop diagram with a soft photon (see the first diagram in
Fig.~\ref{figselfenergy}). This part arises from scatterings where
the electrons exchange a soft photon screened at the scale of the Debye
mass.

In order to verify the Ward identity up to a given order in the coupling
constant, all processes that contribute up to that order have to be
included (in particular, not just those processes that contribute to
transport). Therefore, we compute in this section the complete
contribution to the thermal width to order $e^4T\ln(1/e)$. In
Sec.~\ref{de} we show how the scale $e^4T\ln(1/e)$ actually arises in the
field theory calculation of the conductivity, from both soft photon and
soft fermion mediated scattering processes. It turns out that only the
soft fermion contribution to the thermal width appears explicitly. The
processes in which a soft photon, screened at the scale of the Debye mass,
is exchanged contribute not through the thermal width but in an indirect
way, through the rungs in the ladder diagrams.

The thermal width of an on-shell electron is given by\footnote{The same 
result is obtained if one uses 
$\tw_{\pv}=-2\im\,\Sigma^{R}_{-}(\po=-p,\pv)$.}
\be
 \Gamma_{\pv} = -2\im\, \Sigma^{R}_{+}(p^0=p,\vecp).
\ee
The one-loop fermion self-energy reads
\be
 \Sigma(P) = -e^{2}\si_K \gamma^{\nu}S(P+K)\gamma^{\mu}D_{\mu\nu}(K).
\ee
The Matsubara sum is easily performed using spectral representations. For 
the photon we work in the Coulomb gauge and the photon propagator reads
\be
 D_{\mu\nu}(p^0,\pv)=-\frac{1}{\pv^2}P^L_{\mu\nu}-\int_{-\infty}^{\infty} 
 \frac{d\omega}{2\pi}\, \frac{ \rho_{\mu\nu}(\om,\pv) }{p^0-\omega},
\ee
with
\be
 \rho_{\mu\nu}(\om,\pv) = \rho_{T}(\omega,\pv)P^T_{\mu\nu}(\pvuni) + 
 \rho_{L}(\omega,\pv) P^L_{\mu\nu},
\ee
and $P^{T}_{ij}(\pvuni)=\delta_{ij}-\puni_{i}\puni_{j}$, 
$P^{T}_{0\nu}=\cP^{T}_{\mu0}=0$ and 
$P^{L}_{\mu\nu}=\delta_{\mu0}\delta_{\nu0}$. We find for the imaginary 
part of the retarded on-shell self-energy,
\be
 \im\,\Sigma^{R}(p,\vecp) = -\frac{e^2}{2}\int_{\veck,\om} 
 \left[n_B(\om)+n_F(p+\om)\right] \gamma^\nu \rho_F(p+\om,\vecr) 
\gamma^\mu \rho_{\mu\nu}(\om,\veck),
\ee 
with $\vecr=\vecp+\veck$, $n_B(\om)=1/[\exp(\om/T)-1]$ is the Bose 
distribution, and 
\be
 \rho_F(\om,\kv) = 
 \rho_{+}(\om,\kv)h_{+}(\kvuni)+\rho_{-}(\om,\kv)h_{-}(\kvuni).
\ee 
With the help of the following useful relations,
\bea
 \nn && \hm 
 h_{\pm}(\pvuni)h_{\pm}(\pvuni)=0, 
 \;\;\;\;\;\;\;\;\;\;\;\;\;\;
 h_{\pm}(\pvuni)h_{\mp}(\pvuni) = \gamma^{0}h_{\mp}(\pvuni) 
 = h_{\pm}(\pvuni)\gamma^{0}, 
 \\ && \hm 
 h_{\pm}(\pvuni)\gz h_{\mp}(\pvuni)=0,
 \;\;\;\;\;\;\;\;\;\;\;
 h_{\pm}(\pvuni)\gi h_{\mp}(\pvuni) = (\pm\puni^{i}-\gi\gz)h_{\mp}(\pvuni),
 \\ \nn && \hm 
 h_{\pm}(\pvuni)\gz h_{\pm}(\pvuni)=h_{\pm}(\pvuni), 
\eea
and
\be
 \label{eqPTPL}
 \gamma^\mu h_\pm(\pvuni) \gamma^\nu P^L_{\mu\nu} =  h_\mp(\pvuni),
 \;\;\;\;\;\;\;\;\;
 \gamma^\mu h_\pm(\pvuni) \gamma^\nu P^T_{\mu\nu}(\rvuni) =  
\gamma^0 \mp \pvuni\cdot\rvuni \, \gv\cdot\rvuni,
\ee
the (exact) result for the one-loop width is
\bea   \label{a}
 \tw_{\pv} = &&\hm e^{2} \int_{\veck,\om} 
 \left[n_{F}(p+\omega)+n_{B}(\omega)\right] \nn \\
 &&\hm\nn
 \times \left(\, \rho_{T}(\omega,\kv) \left[
 \rho_{+}(p+\omega,\rv)(1-\pvuni\cdot\kvuni\, \kvuni\cdot\rvuni) + 
 \rho_{-}(p+\omega,\rv)(1+\pvuni\cdot\kvuni\, \kvuni\cdot\rvuni) 
 \right] \right. \\ 
 &&\hm \left. 
 + \frac{1}{2} \rho_{L}(\omega,\kv) 
 \left[\rho_{+}(p+\omega,\rv)(1+\pvuni\cdot\rvuni) 
 + \rho_{-}(p+\omega,\rv)(1-\pvuni\cdot\rvuni) 
 \right]\right).
\eea

\begin{figure}[t]  
 \centerline{
 \epsfig{figure=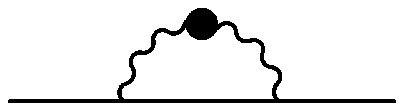,height=2cm}
 \epsfig{figure=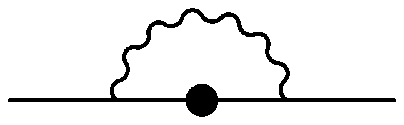,height=2cm} 
 }
 \centerline{ (sp) \hspace{4.5cm} (sf) }
 \caption{Contributions to the thermal width of a hard on-shell fermion
 with a soft photon (sp) and a soft fermion (sf).}
 \label{figselfenergy}
\end{figure}

There are two contributions of order $e^4T\ln(1/e)$, arising when either
the photon or the fermion carries soft momentum, 
$\Gamma_\pv = \Gamma_\pv^{\rm (sp)} + \Gamma_\pv^{\rm (sf)}$ (see
Fig.~\ref{figselfenergy}). We first specialize to the case that the photon
is soft, $k\ll p$. In this case the momentum of the fermion inside the
loop is hard and its spectral density can be taken as the free one,
$\rho_\pm^{\rm free}(\om,\pv)=2\pi\delta(\om\mp p)$. Since we consider
$\po=p$, $\rho_{-}$ does not contribute and we have
\bea  
 \tw_{\pv}^{\rm (sp)} =  &&\hm e^{2} \int_{\veck,\om} 
 \left[n_{F}(p+\omega)+n_{B}(\omega)\right] 
 \rho_{+}^{\rm free}(p+\omega,\rv) 
  \nn \\ &&\hm \times 
 \left[ 
 \aste\rho_{T}(\omega,\kv) (1-\pvuni\cdot\kvuni\, \kvuni\cdot\rvuni) 
 + \frac{1}{2} \aste\rho_{L}(\omega,\kv) (1+\pvuni\cdot\rvuni) 
 \right].
 \label{eqsp}
\eea
The angular integration can be performed with the fermionic spectral 
function,
\be
 \rho_{+}^{\rm free}(p+\omega,\rv) = 2\pi\delta(p+\omega-r) \to
 2\pi\frac{p+\omega}{pk}\,\delta(z-z_{0})\theta(k^{2}-\om^{2}), 
 \;\;\;
 z_{0} = \frac{\om}{k} + \frac{\omega^2-k^2}{2pk},
\ee
where $z$ is the cosine of the angle between $\kv$ and $\pv$. We 
find for the contribution with the soft photon
\bea  
 \tw^{\rm{(sp)}}_{\pv} = &&\hm \frac{\al}{2 p^{2}}
 \int_{\lmin}^{\lmax} \!\!\!\!dk\,k \int_{-k}^{k}\frac{d\om}{2\pi}\, 
 \left[n_{F}(p+\omega)+n_{B}(\omega)\right]  
 \nn \\ &&\hm  \label{twsp}
 \times \left( \aste\rho_{T}(\om,k)\frac{k^2-\om^2}{k^2} 
 \left[(\om+2p)^2+k^2\right] 
 + \aste\rho_{L}(\omega,k)\left[(\om+2p)^2-k^2\right] \right),
\eea
with $\al=e^2/4\pi$. The integral over the momentum $k$ has been restricted 
between $\lmin$, a lower cutoff to avoid the logarithmic singular behaviour, 
and $\lmax$ (with $eT\ll\lmax\ll T$ \cite{braaten}), 
so that the approximation of soft photon momentum is valid. 
In order to find contributions up to $e^4T\ln(1/e)$, 
we define $x=\om/k$ and expand in powers of $k/p$,
\bea  
 \Gamma_\vecp^{\rm (sp)} = 2\al T \int_{\lmin}^{\lmax} \!\!\!\!dk\,k
   &&\hm 
 \int_{-k}^{k} \frac{d\om}{2\pi}\, \frac{1}{\om}
 \left[ \aste\rho_{T}(\om,k)\left( V_{T}^{(0)}(x) 
 + V_{T}^{(1)}(x)\frac{k}{p}
 + V_{T}^{(2)}(x)\frac{k^{2}}{2p^{2}} + \ldots\right) \right.
 \nn \\
 \label{spexp}
 &&\hm\left. 
  +\, \aste\rho_{L}(\omega,k)\left(V_{L}^{(0)}(x) 
  + V_{L}^{(1)}(x)\frac{k}{p}
  + V_{L}^{(2)}(x)\frac{k^{2}}{2p^{2}}+\ldots\right)\right],
\eea
with
\bea
 V_{L}^{(0)}(x) = &&\hm 1,      \hspace{5.3cm}
 V_{T}^{(0)}(x) = 1-x^{2},
 \\   
 V_{L}^{(1)}(x) = &&\hm \frac{1}{2} x(2-\beta p\,[1-2n_{F}(p)]), 
 \;\;\;\;\;\;\;\;\;\;\,
 V_{T}^{(1)}(x) = (1-x^{2})V_{L}^{(1)}(x),  
 \\
 V_{L}^{(2)}(x) = &&\hm -\frac{1}{2}(1-x^{2}) - \beta 
 p\,x^{2}[1-2n_{F}(p)] 
 + \frac{1}{6}\beta^{2}p^{2}x^{2}[1+12\beta n_{F}'(p)],     
 \\
 V_{T}^{(2)}(x) = &&\hm (1-x^2)[1+V_L(x)], 
\eea
where $\beta=1/T$. The integrals over $\om$ can be performed using sum 
rules (see Appendix 
\ref{appsumrules}). It is convenient to split the range of integration 
between $\lmin$ and the Debye mass $m_{D}\sim eT$, and between $m_{D}$ and 
$\lmax$, such that the residues and dispersion relations can be 
approximated in both ranges. For the leading-order term $V_{T/L}^{(0)}$, 
the dominant contribution comes from the lower part of the integral and 
from transverse photons only. We recover the logarithmic singular behaviour 
\cite{smilga,blaizot}
\be  \label{leading}
 \Gamma^{\rm (sp,lo)}_{\pv} = 2\al T\lmg.
\ee
With the help of the sum rules one can show that subleading corrections 
[to $e^2\ln(m_{D}/\lmin)$] do not lead to $e^4\ln(1/e)$ behaviour.
The next contribution, from $V_{T/L}^{(1)}$, vanishes because it is odd in 
$\om$. Therefore the next-to-leading order contribution to the 
thermal width comes from $V_{T/L}^{(2)}$. This contribution is finite and 
$\lmin$ can be taken to zero.
In this case sum rules show that the dominant contribution  arises from 
momentum $m_D \ll k \ll \lmax$. We may take $\lmax\sim T$, since we are only 
interested in the coefficient of the logarithmic term \cite{braaten}. 
Performing the integral over $\om$ with the sum rules 
collected in Appendix \ref{appsumrules} we arrive at
\be
 \label{twspnlo}
 \Gamma^{\rm (sp,nlo)}_{\pv} = 
 \frac{\alpha\, m_D^2\ln(1/e)}{2p}
 \left[-1+2n_{F}(p)+\frac{p}{6T}+ 2p\,n_{F}'(p) \right],
\ee
with $m_D^2 = e^2T^2/3$. We note here that the leading logarithmic terms
in the sum rules [see Eq.\ (\ref{eqsumphoton})] cancel exactly. We also
note that this contribution is negative for momentum $p\lesssim 6T$.
Higher-order terms in the expansion in $k/p$ of Eq.~(\ref{twsp}) yield
contributions parametrically suppressed with respect to $e^4T\ln(1/e)$.

Now we turn to the contribution when the fermion is soft. Since in this
case the momentum of the photon is hard, only the free transverse photon
contributes. Making a change of variables ($p+\om \to -\om$, 
$\pv+\kv\to-\kv$) such that the fermion carries
the soft momentum $\veck$, we get
\bea   
 \nn
 \Gamma^{\rm(sf)}_{\pv} = &&\hspace*{-0.6cm}e^{2}\int_{\veck,\om}
 \left[n_{F}(\omega) + n_{B}(p+\om)\right] \rho_{T}^{\rm free}(p+\omega,\rv)
 \\ \label{widthsofte}  &&\hspace*{-0.6cm}
 \times\left[
 \aste\rho_{+}(\omega,\kv)(1-\pvuni\cdot\rvuni\,\kvuni\cdot\rvuni)+
 \aste\rho_{-}(\omega,\kv)(1+\pvuni\cdot\rvuni\,\kvuni\cdot\rvuni)
 \right].
\eea
The angular integration can be performed using the photon spectral 
function,
\be  
 \label{rsf}
 \rho^{\rm free}_{T}(p+\omega,\rv) = 
 2\pi\sgn(p+\om)\delta((p+\om)^{2}-r^{2}) 
 \longrightarrow 
 \frac{2\pi}{2pk}\delta(z-z_0)\theta(k^{2}-\om^{2}). 
\ee
As a result we get
\be
 \Gamma^{\rm(sf)}_{\pv} = \frac{\al}{4 p^2} 
 \int_{0}^{\lmax}\!\!\!\! dk\int_{-k}^{k}\frac{d\omega}{2\pi} 
 \left[n_{F}(\om)+n_{B}(p+\omega)\right]
 \left[ V_+{}^*\!\rho_{+}(\omega,k) + V_-{}^*\!\rho_{-}(\omega,k)\right],
\ee
with
\be
 V_\pm = \frac{k\mp\om}{(p+\om)^2}\left[4p^3 + 2(3\om \mp k)p^2 
 + 4\om^2p +(\om\pm k)(k^2+\om^2)\right].
\ee
Since this integral is well-defined for $k\to 0$, one may safely take 
$\Lambda_{min}=0$.
We proceed as in the case of the soft photon and expand in $k/p$ after 
introducing $x=\om/k$. Using the sum rules for HTL fermion spectral 
functions it is easy to see that the leading-order contribution to the 
thermal width comes from the first term in the expansion,
\be
 \Gamma^{\rm (sf,lo)}_{\pv} = \al\frac{1+2n_{B}(p)}{2 p} 
 \int_{0}^{\lmax} \!\!\!\!dk\, k \int_{-k}^{k} \frac{d\omega}{2\pi}
 \left[
 \left(1-\frac{\om}{k}\right) {}^*\!\rho_{+}(\omega,k) + 
 \left(1+\frac{\om}{k}\right) {}^*\!\rho_{-}(\omega,k)
 \right],
\ee
when the soft fermion momentum $k$ lies in the range $m_f \ll k \ll \lmax$. 
Here $m_f^2=e^2T^2/8$ is the fermion thermal mass squared. With the help 
of the sum rules listed in Appendix \ref{appsumrules} and using that to 
leading-logarithmic accuracy $ \lmax\sim T$, the result is 
\be \label{twsf}
 \Gamma^{\rm(sf,lo)}_{\pv} = 
 \frac{\alpha\,m_f^2\ln(1/e)}{p}\left[1+2n_{B}(p)\right].
\ee
As in the case of the soft photon, the leading logarithmic terms in the
sum rules [see Eq.~(\ref{eqsumfermion})] cancel exactly. This result, of
course, agrees with Ref.~\cite{valle}.


\section{Ward identity}
\label{ward}

The Ward identity for the electron-photon vertex in QED is
\be
 Q_{\mu}\Gamma^{\mu}(P+Q,P) = S^{-1}(P)-S^{-1}(P+Q). 
\ee
As shown in Sec.~\ref{ec}, the effective vertex appearing in 
the expression for the electrical conductivity is given by the following 
analytic continuation, 
\be  \label{ac}
 i\om_{p}+i\om_{q}\longrightarrow\po+\qo+i0^{+}, 
 \hp \hp i\om_{p}\longrightarrow\po-i0^{+},
\ee
with $\qv=0$.
Thus the Ward identity reads
\be
 \qo\tw^{0}(\po+\qo+i0^{+},\po-i0^{+};\pv) = 
  S_A^{-1}(P)- S_R^{-1} (P+Q) 
 = \qo\gz + \Sigma^{A}(P)-\Sigma^{R}(P+Q).
\ee
In order to make this a scalar equation, we may contract it with positive- 
or negative-energy spinors and find
\be
 \label{eqWD}
 \qo D_\pm^0(\po + \qo, \po; \pv) = \qo + i\tw_\pv,
\ee
where we used Eq.~(\ref{eqsigmaRA}) for the self-energies, 
definitions (\ref{eqDplus}, \ref{eqDmin}) for $D_\pm^0$, as well as
\bea
 \label{equu}
 \bar{u}_{\lambda}(\pvuni)\gamma^0 u_{\lambda^\prime}(\pvuni) 
 = \delta_{\lambda\lambda^\prime},
 && \;\;\;\;
 \bar{u}_{\lambda}(\pvuni)\gi u_{\lambda^\prime}(\pvuni) 
 = \puni^i \delta_{\lambda\lambda^\prime}, \\
 \bar{v}_{\lambda}(\pvuni)\gamma^0 v_{\lambda^\prime}(\pvuni) = 
 \delta_{\lambda\lambda^\prime},
 && \;\;\;\;
 \bar{v}_{\lambda}(\pvuni)\gi v_{\lambda^\prime}(\pvuni)   
 = -\puni^i \delta_{\lambda\lambda^\prime}.
\eea
We emphasize that Eq.~(\ref{eqWD}) is only valid in the special 
kinematical regime relevant 
for the conductivity, i.e.\ $\qo\to 0$ and $\po\simeq\pm p$. To make 
this explicit, we define the quantity
\be
 \label{eqdefD}
 {\mathfrak{D}}(\pv) \equiv \lim_{\qo\to 0} \qo
 D_\pm^0(\pm p+\qo, \pm p;\pv). 
\ee
The Ward identity is then simply 
\be
 \label{eqWI}
 \mathfrak{D}(\pv) = i\tw_\pv.
\ee 
To verify that the integral equation $\Gamma^0 = \gamma^0 + \Gamma^0_{\rm
HTL} + \Gamma^0_{\rm ladder}$ (see Fig.\ \ref{figie}) is consistent with
the Ward identity, we choose to continue with $\po=p$ and contract 
the integral equation with 
positive-energy spinors
$\bar{u}_{\lambda}(\pvuni)\dots u_{\lambda}(\pvuni)$, multiply it with
$\qo$ and take the limit $\qo\to 0$, $\po=p$. The tree level contribution
then vanishes. The two remaining parts on the right-hand-side should
give $i$ times the thermal width,
$\Gamma_\pv = \Gamma_\pv^{\rm (sp)} +\Gamma_\pv^{\rm (sf)}$.

We start with the term $\tw_{\rm HTL}^{0}$ in the integral equation, 
\be
 \tw_{\rm HTL}^{0}(P+Q,P) = e^{2}\si_{K}
 \gamma^{\mu}\aste S(K+Q)\aste\tw^{0}(K+Q,K)
 \aste S(K)\gamma^{\nu}D_{\mu\nu}(P-K),
\ee
where again $Q=(i\om_q,\vecnul)$. 
We can use the (euclidean) Ward identity satisfied by the HTL vertex
\be
 \aste S(K+Q)\aste\tw^{0}(K+Q,K)\aste S(K) =
 \frac{1}{i\om_{q}}\left[\aste S(K+Q)-\aste S(K)\right],
\ee
to simplify the expression,
\be
 \tw_{\rm HTL}^{0}(P+Q,P) = \frac{e^{2}}{i\om_{q}}\si_{K}\gamma^{\mu}
 \left[\aste S(K+Q)-\aste S(K)\right]\gamma^{\nu}D_{\mu\nu}(P-K).
\ee
Since $P$ is hard and $K$ soft, we need only to consider free transverse
photons. Using spectral representations for the propagators it is
straightforward to do the sum over the Matsubara frequencies, arriving at
\bea
 \tw_{\rm HTL}^{0}(P+Q,P) =
 -\frac{e^{2}}{i\om_{q}}\int_{\kv,\om,\om'}
 \left[n_{F}(\om)+n_{B}(\om')\right] 
 \gamma^{\mu}\aste\rho_F(\om,\kv)\gamma^{\nu} 
 P^{T}_{\mu\nu}(\rvuni)\rho_{T}^{\rm free}(\om',\rvuni)
 \nonumber &&\\
 \times \left( \frac{1}{i\om_{p}+i\om_{q}+\om-\om'} - 
 \frac{1}{i\om_{p}+\om-\om'} \right).&&
\eea
Now we can do the analytic continuation (\ref{ac}), 
choose $\po=p$, multiply with $\qo$ and take it to zero, to arrive at
\be
 \lim_{\qo\to 0} \qo\tw_{\rm HTL}^{0} = 
 ie^{2}\int_{\kv,\om,\om'}
 \left[n_{F}(\om)+n_{B}(\om')\right]
 \gamma^{\mu}\aste\rho_F(\om,\rv)\gamma^{\nu}
 P^{T}_{\mu\nu}(\rvuni)\rho_{T}^{\rm free}(\om',\rvuni)
 2\pi\delta(p+\om-\om').
\ee
In the limit $\qo\to 0$ the real part of the vertex multiplied with $\qo$ 
vanishes. The remaining part is purely imaginary, as 
required by the Ward identity.
After performing the integral over $\om'$ and using (\ref{eqPTPL}) to do 
the algebra, we contract with the positive-energy spinors and arrive at 
\bea 
 \mathfrak{D}_{\rm HTL}(\pv) = &&\hm ie^{2}\int_{\kv,\om} 
 \left[ n_{B}(p+\om) + n_{F}(\om)\right] \rho_{T}^{\rm free}(p+\om,\rv)        
 \nonumber \\ &&\hm
 \times 
 \left[\aste\rho_{+}(\om,\kv)(1-\pvuni\cdot\rvuni\,\kvuni\cdot\rvuni) +
 \aste\rho_{-}(\om,\kv)(1+\pvuni\cdot\rvuni\,\kvuni\cdot\rvuni)\right].
 \label{eqDhtl}
\eea
The right-hand-side of Eq.\ (\ref{eqDhtl}) is precisely $i$ times the 
contribution from the soft fermion to the thermal width $\Gamma_\pv^{\rm 
(sf)}$, see Eq.~(\ref{widthsofte}).

Now we turn to the remaining contribution $\tw_{\rm ladder}^{0}$. We have
\be
 \tw_{\rm ladder}^{0}(P+Q,P) = e^{2}\si_{K}\gamma^{\mu}S(P+K+Q)
 \tw^{0}(P+K+Q,P+K)S(P+K)\gamma^{\nu}\aste D_{\mu\nu}(K).
\ee
We can do the sum of Matsubara frequencies using the contour of 
Ref.~\cite{valle} and perform the analytic continuation (\ref{ac}). We 
choose 
again $\po=p$, multiply with $\qo$ and take it to zero. This gives
\bea
 \lim_{\qo\to 0} \qo\Gamma^0_{\rm ladder} = &&\hm
 e^{2}\di_{\kv,\om}\left[n_{B}(\om)+n_{F}(p+\om)\right]
 \Delta^{R}_{+}(p+\om,\rv) \Delta^{A}_{+}(p+\om,\rv) 
 \nn \\ &&\hm \times 
 \gamma^\mu h_+(\rvuni)\qo\Gamma^0(p+\om+i0^+,p+\om-i0^+;\rv)
 h_+(\rvuni)\gamma^\nu \aste\rho_{\mu\nu}(\om,\kv).
\eea
Here we used that in the pinching-pole limit (\ref{eqpp}) with 
$\po=p$ only the positive-energy propagators contribute. 
A convenient way to proceed is to realize that the full vertex 
$\Gamma^\mu$ is linear in the $\gamma$-matrices.\footnote{This can be seen 
by decomposing the vertex in the 16 basis elements $1$, $\gamma^\mu$, 
$\gamma_5$, $\gamma_5\gamma^\mu$, 
$\sigma^{\mu\nu}=i[\gamma^\mu,\gamma^\nu]/2$.
The integral equations for the coefficients that are not linear in the 
$\gamma$-matrices decouple.} 
Since the vertex then conserves helicity (see e.g.\ Eq.\ (\ref{equu})) we 
may use
\be
 \label{eqlinear}
 h_\pm(\rvuni) \Gamma^\mu(\pm p+\om+i0^+, \pm p+\om-i0^+;\rv) 
h_\pm(\rvuni)
 = h_\pm(\rvuni) D^\mu_\pm(\pm p+\om,\pm p+\om;\rv).
\ee
Using then again Eq.~(\ref{eqPTPL}) and contracting with the 
positive-energy spinors gives
\bea
 \mathfrak{D}_{\rm ladder}(\pv) = &&\hm
 e^{2}\di_{\kv,\om}\left[n_{B}(\om)+n_{F}(p+\om)\right]
 \Delta^{R}_{+}(p+\om,\rv) \Delta^{A}_{+}(p+\om,\rv)  \,\mathfrak{D}(\rv)
 \nn \\ &&\hm \times 
 \left[
 \aste\rho_{T}(\omega,\kv) (1-\pvuni\cdot\kvuni\, \kvuni\cdot\rvuni)
 + \frac{1}{2} \aste\rho_{L}(\omega,\kv) (1+\pvuni\cdot\rvuni)
 \right].
 \label{eqDladder}
\eea
In the pinching-pole limit (\ref{eqpp}) the product of the propagators is 
proportional to $1/\Gamma_\pv$. It is then easy to see that our integral  
equation 
is consistent with the Ward identity (\ref{eqWI}). If we use the 
Ward identity itself explicitly we can write 
\be
 \Delta^{R}_{+}(p+\om,\rv) \Delta^{A}_{+}(p+\om,\rv)  \,\mathfrak{D}(\rv)
 = 2\pi i \delta(p+\om-r).
\ee
Eq.~(\ref{eqDladder}) then indeed yields precisely $i$ times the
contribution to the thermal width from the soft photon 
$\Gamma_\pv^{\rm (sp)}$ in
Eq.~(\ref{eqsp}). We conclude that with both $\Gamma_{\rm HTL}^0$ and
$\Gamma_{\rm ladder}^0$ the Ward identity is satisfied.
We point out that the Ward identity relates the diagrams in the vertex 
equation to those contributing to the electron self-energy 
exactly, without doing any approximation.


\section{Integral equation for the spatial part of the vertex}
\label{de}

In the previous section we verified that the modified vertex equation
summing the ladders is consistent with the Ward identity. Now we turn to
the spatial part of the vertex equation, which appears in the expression
for the transport coefficient.

First we consider the contribution of the new diagram. It has an imaginary
part which behaves as $\sim 1/\qo$ in the limit that $\qo\to 0$, due to
the structure of the HTL vertex. However, the conductivity only depends on
the real part of the vertex, so we focus on the real part only. Since it
is a modification of the tree level vertex (defined to be $\gamma^{\mu}$),
in order to be relevant for the calculation of the electrical
conductivity, it should be at least of order 1.  The interaction vertices
in the diagram give a factor $e^2$.  One could expect that pinching poles
might be present and compensate for the explicit powers of the coupling
constant;  however it turns out that the frequency of the fermion
propagators is always below the light-cone and therefore the poles of the
HTL electron propagator, which lie above the light-cone, can never be
reached. The conclusion is therefore that in the limit $\qo\to 0$ the real
part of the new diagram is finite and smaller than the tree level vertex.  
This is shown explicitly in Appendix \ref{spatial}.  In fact, explicit
power counting shows that it is suppressed by three powers of the
coupling.

It only remains to compute the contribution from the diagram with the soft
rung. This was, in leading-logarithmic order, done in Ref.~\cite{valle}.
Here we derive the leading-log equation for the effective vertex keeping
the identification with the explicit expression of the self-energy
completely general, which allows us to correct a small error in the
derivation of Ref.~\cite{valle}.

After doing the Matsubara frequency sum, the diagram reads
\bea
 &&\hm 
 \Gamma^i_{\rm ladder}(p+\qo+i 0^+, p-i 0^+;\pv) = 
 e^2 \di_{\kv,\om} \left[n_{B}(\om)+n_{F}(p+\om+\qo)\right] 
 \Delta^{R}_{+}(p+\om+\qo,\rv)
 \nn \\ && \times 
 \Delta^{A}_{+}(p+\om,\rv)
 \gamma^{\mu} h_+(\rvuni) \Gamma^i(p+\om+\qo+i0^+,p+\om-i0^+;\rv) 
 h_+(\rvuni) \gamma^{\nu} \aste\rho_{\mu\nu}(\om,\kv),
\eea
where we recall that $\rv=\pv+\kv$.
We choose to take $\po=p$ and since $\qo$ will be taken to zero, only 
positive-energy propagators contribute. To proceed, we use 
property (\ref{eqlinear}) and Eq.\ (\ref{eqPTPL}) to do the 
algebra and contract with positive-energy spinors
$\bar{u}_{\lambda}(\pvuni)\ldots u_{\lambda}(\pvuni)$.\footnote{We remind
that one could as well contract with $\bar{v}_{\lambda}(\pvuni)\ldots
v_{\lambda}(\pvuni)$ and use $\po=-p$. Since $\cD_{+}(p) = -\cD_{-}(p)$ it
does not matter which one is used.} 
Since in the pinching-pole limit everything is real except the vertex itself, 
the real and the imaginary parts of the integral equation 
decouple. Recalling the property $\cD^{i}_{+}(\pv) = \puni^{i}\cD(p)$, we 
can multiply the real part of the integral equation with $\puni^i$ and 
find, after doing the angular integral,
\bea  
 \cD(p) = 1 +  &&\hm \nn  \frac{\al}{2 p^{2}} 
 \int_{\lmin}^{\lmax}\!\!\!\!\!\! dk\,k \int_{-k}^{k}\frac{d\om}{2\pi}\,
 \left[n_{B}(\om)+n_{F}(p+\om)\right] 
 \left\{ \pvuni\cdot\rvuni\frac{\cD(r)}{\tw_{\rv}}\Big|_{z=z_{0}} 
 \right\}  \\  
 &&\hm \times 
 \left[\aste\rho_{T}(\om,k)
 \frac{k^{2}-\om^{2}}{k^{2}}\left[(\om+2p)^{2}+k^{2}\right]
 +\aste\rho_{L}(\omega,k)\left[(\om+2p)^{2}-k^{2}\right] \right].
 \label{ie0}
\eea
We notice that, save for the factor within braces, the integral is
precisely Eq.~(\ref{twsp}) giving the soft photon contribution
$\tw^{(\rm{sp})}_{\pv}$ to the thermal width. Now we define
\be
 \chi(p)\equiv \frac{\cD(p)}{\tw_{\pv}},
\ee
with $\tw_{\pv} = \tw^{(\rm{sp})}_{\pv} + \tw^{(\rm{sf})}_{\pv}$ 
and get for the integral equation
\bea 
 \nn \hm1 = \tw_\pv^{(\rm{sf})} \chi(p) + 
 \frac{\al}{2 p^{2}} 
 \int_{\lmin}^{\lmax}\!\!\!\!\!\! dk\,k \int_{-k}^{k}\frac{d\om}{2\pi}\,
 \left[n_{B}(\om)+n_{F}(p+\om)\right] 
 \left\{\chi(p)-\pvuni\cdot\rvuni\;\chi(r)\Big|_{z=z_{0}} \right\}        
 &&\\ 
 \times \left[\aste\rho_{T}(\om,k)\frac{k^{2}-\om^{2}}{k^{2}}
 \left[(\om+2p)^{2}+k^{2}\right]
 +\aste\rho_{L}(\omega,k)\left[(\om+2p)^{2}-k^{2}\right]\right]&&\hm.   
 \label{ie} 
\eea
So far we have made no approximation, apart from taking the pinching-pole 
limit. To arrive at the leading-log approximation, we write
$x=\om/k$ and expand in powers of $k/p$. We need to expand the term 
in braces up to second order in $k/p$,
\be \label{chi}
 \chi(p)-\pvuni\cdot\rvuni\;\chi(r)\big|_{z=z_{0}} =
 -x p \chi'(p)\frac{k}{p} + 
 \left[(1-x^{2})\chi(p)-x^{2}p^{2}\chi''(p)\right]\frac{k^{2}}{2p^{2}} 
 + \ldots
\ee
The expansion of the other terms is precisely as in Eq.~(\ref{spexp}). 
To leading order in $k/p$ (which gives the leading-log order) we find
\be
 \label{eqDlo}
 1 = \tw_\pv^{(\rm{sf,lo})}\chi(p) +
 \frac{2\al T}{p^{2}}
 \int_{\lmin}^{\lmax}\hs{-0.3cm}dk\, k^3 \int_{-k}^{k}\frac{d\om}{2\pi}
 \frac{1}{\om}   
 \left[\aste\rho_{T}(\om,k)\tilde V_{T}(\om/k) + 
 \aste\rho_{L}(\omega,k)\tilde V_{L}(\om/k)\right],
\ee
with
\bea
 \tilde V_{L}(x)= &&\hm \frac{1}{2}[(1-x^{2})\chi(p) - p\,x^{2}\left( 
 2-p\beta\left[1-2n_{F}(p)\right]\right)\chi'(p) - p^{2}x^{2}\chi''(p)], \\
 \tilde V_{T}(x)= &&\hm (1-x^{2})\tilde V_{L}(x).
\eea
It is worth noting that although $V_{T/L}^{(1)}$ did not contribute to the 
thermal width, it is required here to get the leading order result,
\be
 \tilde V_{T/L}(x) = 
 \frac{1}{2} 
\left[ (1-x^{2})\chi(p)-x^{2}p^{2}\chi''(p)\right] V_{T/L}^{(0)}(x) - x p 
\chi'(p)V_{T/L}^{(1)}(x). 
\ee
Thus, taking into account the relation between the soft photon rung 
and the soft photon contribution to the thermal width, it is necessary to 
go beyond the contributions $V_{T/L}^{(0)}$ that gives the leading 
logarithmic 
contribution to the thermal width.\footnote{In Ref.~\cite{valle} the term 
$V_{T/L}^{(1)}$ was neglected. This error was luckily cancelled by 
another coming from doing the expansion (\ref{chi}) with just the leading 
term in $z_{0}$.}
Furthermore, because $V_{T}^{(0)}$ is now multiplied by two additional
powers of $k$, it gives a finite contribution and no dependence on $\lmin$
arises. Finally, $V_{T/L}^{(2)}$, which led to $\Gamma_\pv^{(\rm
sp,nlo)}$, turns out to be irrelevant since it appears only in subleading
terms.

Using sum rules it is easy to see that the dominant contribution comes 
from momenta $m_D<k<\lmax$, and again to leading-log 
accuracy we may take $\lmax\sim T$. Performing the integral over $\om$ 
with the help of the sum rules and using Eq.~({\ref{twsf}}) for 
$\Gamma_\pv^{\rm (sf,lo)}$, we arrive at~\cite{amy-tc,valle}
\bea
 1 = &&\hm \frac{\alpha\, m_{f}^{2}\ln(1/e)}{p} \left[1+2n_{B}(p)\right] 
 \chi(p) \nn \\
 &&\hm + \frac{\alpha\, m_{D}^{2}\ln(1/e)}{p} \frac{T}{p}
 \left[ \chi(p) - 
 \left(1-\frac{p}{2T}\,[1-2n_{F}(p)]\right) p\,\chi'(p)
 -\frac{1}{2}p^{2}\chi''(p) \right].
\eea
Again the leading-logarithmic terms in the sum rules [see 
Eq.~(\ref{eqsumphoton})] cancel exactly.
The electrical conductivity is then given by 
\be
 \sigma=-\frac{4e^{2}}{3}\int_{\pv}n_{F}'(p)\chi(p).
\ee
The parametrical behaviour of the conductivity can be made explicit by 
writing
\be
 \chi(p) = \frac{T}{\alpha\,m_D^2 \ln(1/e)}\,\phi(p/T),
\ee
such that
\be
 \sigma = C\frac{T}{e^{2}\ln(1/e)}, \;\;\;\;\;\;\;\;\;\;\;\;\;\;
 C = \frac{2}{\pi}\int^{\infty}_{0} dy\, y^2\frac{1}{\cosh^2(y/2)}\phi(y).
\ee
The dimensionless function $\phi(y)$ obeys the differential equation
\be
1 = \left[ \frac{3\coth(y/2)}{8y}+\frac{1}{y^2}\right]\phi(y) 
+ \left[\frac{1}{2}\tanh(y/2)-\frac{1}{y}\right]\phi'(y) 
-\frac{1}{2}\,\phi''(y).
\ee
To obtain the final result for the conductivity, the differential equation 
should be solved or, alternatively, an equivalent variational 
problem as was done in Ref.~\cite{amy-tc}, where the value $C =15.6964$ 
was obtained.


\section{Conclusions}
\label{seccon}

The computation of the electrical conductivity in hot QED at
leading-logarithmic order requires the summation of an infinite series of
ladder diagrams as well the inclusion of a thermal width for hard on-shell
fermions. We studied the Ward identity for the effective photon-electron
vertex summing these diagrams. In order to match soft fermionic
contributions to the thermal width of order $e^4 T\ln(1/e)$, we found that
a new diagram has be included in the integral equation for the vertex.
This diagram contains a hard photon rung and soft fermion lines 
as well as the associated HTL vertex.

A consequence of the Ward identity is that in the kinematical region
relevant for transport coefficients (external frequency $\qo\to 0$ and
external momentum $\qv=0$), the imaginary part of the temporal component
of the photon-electron vertex $\tw^{0}(P+Q,P)$ is singular $\sim
\Gamma_\pv/\qo$, with $\Gamma_\pv$ the thermal width for hard fermions.  
The real part is finite when $\qo\to 0$ and therefore subdominant.
Similarly the imaginary part of the spatial vertex $\tw^{i}(P+Q,P)$ is
singular. However, in the expression for the electrical conductivity only
the real part of the effective photon-electron vertex appears. We found
that the real part of the new diagram is, in the kinematical regime of
interest, suppressed by three powers of the coupling constant with respect
to the tree-level vertex. Therefore it does not contribute to the
electrical conductivity at leading logarithmic order.  For the same reason
we expect it will also not contribute at full leading order.

The thermal width receives contributions of order $e^{4}T\ln(1/e)$ from
diagrams involving either a soft photon or a soft fermion. Only the
contribution from soft fermions appears explicitly in the expression for
the conductivity to leading-log order. We have verified that the inverse
relaxation time from the Boltzmann equation in the relaxation-time
approximation from those contributions to the collision term where a
fermion is exchanged, i.e.\ diagrams $D$ (fermion annihilation) and $E$
(Compton scattering) in Ref.~\cite{amy-tc}, agrees precisely with the
result (\ref{twsf}). On the other hand, processes contributing to the
thermal width which involve soft photon exchange (i.e.\ Coulomb
scattering) appear in the expression for the electrical conductivity only
indirectly, through the rungs in the ladder diagrams.

For other transport coefficients, such as the shear viscosity, the soft
fermionic contribution to the thermal width contributes as well.  
Therefore it seems that an additional diagram similar to $\tw^{\mu}_{\rm
HTL}$ in our vertex equation will be necessary; the analog of the HTL
vertex in QED but with two fermion lines and one insertion of the operator
$\pi_{ij}$, the traceless spatial part of the energy momentum tensor.  
However, as is the case for the electrical conductivity, this will
probably not affect the leading-log differential equation for the
effective vertex.

Finally, to go beyond the leading-log approximation requires the inclusion
of all contributions to the thermal width that are of order $e^4T$. The
Ward identity may be a useful tool that can help in verifying what type of
diagrams contribute to the conductivity in this case.


\vspace{0.5cm}
\noindent
{\bf Acknowledgements}\\ 
We gratefully acknowledge useful discussions with E.\ Braaten, 
particularly  concerning the Ward identity.
J.~M.~M.~R.\ thanks M.~A.~Valle Basagoiti for helpful conversations.
G.~A.\ is supported by the Ohio State University through a Postdoctoral
Fellowship and by the U.~S.\ Department of Energy under Contract No.\
DE-FG02-01ER41190. 
J.~M.~M.~R.\ is supported by a Postdoctoral Fellowship from the Basque 
Government. This work has been supported in part by the Spanish Science
Ministry under Grants AEN99-0315 and FPA 2002-02037 and by the University
of the Basque Country under Grant 063.310-EB187/98.

\appendix


\section{Sum rules}
\label{appsumrules}

The evaluation of integrals over the Landau damping contribution in HTL 
spectral functions can be conveniently carried out using sum rules 
\cite{LeBellac}. In this Appendix we collect a list of useful results. 

We start with HTL photon spectral functions. We define
\be
 I_n^{T/L}(k) = \int_{-\infty}^{\infty}\frac{d\om}{2\pi}\,
 \om^{2n-1}\,\aste\rho_{T/L}(\om,k).
\ee
The first few sum rules are
\be
 \parbox{2in}{\bean
      I_{0}^{T}(k)&&\hm = \frac{1}{k^{2}}, \\ 
      I_{0}^{L}(k)&&\hm = \frac{m_{D}^{2}}{k^{2}(k^{2}+m_{D}^{2})}, \eean}
 \parbox{2in}{\bean
      I_{1}^{T}(k)&&\hm = 1, \\ 
      I_{1}^{L}(k)&&\hm = \frac{m_{D}^{2}}{3k^{2}}, \eean}
 \parbox{2in}{\bean
      I_{2}^{T}(k)&&\hm = k^{2}+\frac{m_{D}^{2}}{3}, \\ 
      I_{2}^{L}(k)&&\hm = \frac{m_{D}^{2}}{5}+\frac{m_{D}^{4}}{9k^{2}}.
      \eean}
\ee
Because Landau damping contributes below the light-cone only and the pole 
contributions lie inside the light-cone, we find immediately
\be
 J_n^{T/L}(k) \equiv \int_{-k}^{k} \frac{d\om}{2\pi}\, \om^{2n-1} \,\aste\rho_{T/L}(\om,k)
 = I_n^{T/L}(k) - 2Z_{T/L}(k)\om_{T/L}^{2n-1}(k).
\ee
We need these sum rules especially for intermediate momentum $m_D \ll k 
\ll T$. In this case they can be further simplified using 
standard approximations for the residues and dispersion relations 
\cite{LeBellac}. We find
\be      \label{eqsumphoton}
\parbox{3.5in}{
 \bea
 \nn &&J_0^T(k) = 
  \frac{m_{D}^{2}}{4k^{4}}\left[\ln\frac{8k^{2}}{m_{D}^{2}} -1 + 
  \cO\left(\frac{m_D^2}{k^2}\right)\right],\\
 \nn &&J_1^T(k) = 
  \frac{m_{D}^{2}}{4k^{2}}\left[\ln\frac{8k^{2}}{m_{D}^{2}} -3 + 
  \cO\left(\frac{m_D^2}{k^2}\right)\right],\\
 \nn &&J_2^T(k) = 
  \frac{m_{D}^{2}}{4} \left[\ln\frac{8k^{2}}{m_{D}^{2}}-\frac{11}{3} 
  +\cO\left(\frac{m_D^2}{k^2}\right)\right],\eea}
\parbox{2.5in}{
 \bea
  \nn && J_0^L(k) \simeq \frac{m_{D}^{2}}{k^{4}},\\
  \nn && J_1^L(k) \simeq \frac{m_{D}^{2}}{3k^{2}},\\
  \nn && J_2^L(k) \simeq  \frac{m_{D}^{2}}{5}.
 \eea}
\ee
In the case of the longitudinal photons the corrections are exponentially 
suppressed.

For fermionic HTL spectral functions we define
\be
 I_n^\pm(k) = \int_{-\infty}^{\infty} \frac{d\om}{2\pi}\, 
 \om^{n}\,\aste\rho_{\pm}(\om, k),
\ee
and find 
\be
 I_0^\pm(k) =1,\;\;\;\;\;\;\;\; I_1^\pm(k) = \pm k,\;\;\;\;\;\;\; 
 I_2^\pm(k) = k^2+m_f^2.
\ee 
The contribution below the light-cone gives
\be
 J_n^\pm(k) \equiv \int_{-k}^{k} \frac{d\om}{2\pi}\, 
  \om^n \,\aste\rho_{\pm}(\om, k) = 
 I_n^\pm(k) - Z_{\pm}(k)\omega_{\pm}^n(k) 
  - (-1)^n Z_{\mp}(k)\omega_{\mp}^n(k).
\ee
For intermediate momentum $m_f \ll k \ll T$ this yields
\bea  \label{eqsumfermion}
 &&J_0^\pm(k) = \frac{m_{f}^{2}}{2k^{2}}\left[
  \ln\frac{2k^{2}}{m_{f}^{2}} -1 + 
  \cO\left(\frac{m_f^2}{k^2}\right) \right], \nn \\
 &&J_1^\pm(k) = \pm\frac{m_{f}^{2}}{2k}\left[ 
  \ln\frac{2k^{2}}{m_{f}^{2}} -3 + 
  \cO\left(\frac{m_f^2}{k^2}\right) \right], \\
 &&J_2^\pm(k) = \frac{m_{f}^{2}}{2}\left[
  \ln\frac{2k^{2}}{m_{f}^{2}} -3 + 
  \cO\left(\frac{m_f^2}{k^2}\right) \right]. \nn
\eea


\section{Spatial contribution of the new diagram}
\label{spatial}

The new diagram in the integral equation for the effective vertex gives a 
contribution
\be     \label{di}
 \tw^{i}_{\rm HTL}(P+Q,P) = e^{2}\si_{K}\gamma^{\mu}\aste 
S(K+Q)\aste\tw^{i}(K+Q,K)\aste S(K)\gamma^{\nu}D_{\mu\nu}(P-K),
\ee
where the HTL-vertex with vanishing photon momentum is
\be
 \aste\tw^{i}(K+Q,K) \Big|_{\qv=0}  \equiv 
 \aste\tw^{i}(\ko+\qo,\ko;\kv) = 
 A\gz\kuni^{i} + B\gamma^{i} + C\gv\cdot\kvuni\,\kuni^{i},
\ee
with
\bea
 A = &&\hm -\frac{m_{f}^{2}}{k\,\qo}
 \left[Q_{1}\left(\frac{\ko+\qo}{k}\right) - 
 Q_{1}\left(\frac{\ko}{k}\right)\right], \nn \\
 B = &&\hm 1-\frac{m_{f}^{2}}{k\,\qo}
 \left[Q_{2}\left(\frac{\ko+\qo}{k}\right)-Q_{2}\left(\frac{\ko}{k}\right) 
 - Q_{0}\left(\frac{\ko+\qo}{k}\right)+Q_{0}\left(\frac{\ko}{k}\right)
 \right], \\
 C = &&\hm \frac{m_{f}^{2}}{3p\,\qo}
 \left[Q_{2}\left(\frac{\ko+\qo}{k}\right) 
 - Q_{2}\left(\frac{\ko}{k}\right)\right]. \nn  
\eea
Here $Q_{n}(x)$ are Legendre functions of the second kind.
In the limit $\qo\to 0$ the real part of the HTL vertex is regular
whereas the imaginary part (present below the light-cone $k_0^2 < k^2$) 
is singular $\sim 1/\qo$.

\begin{figure}[t]
 \begin{center}
  \scalebox{0.8}[0.8]{\includegraphics*{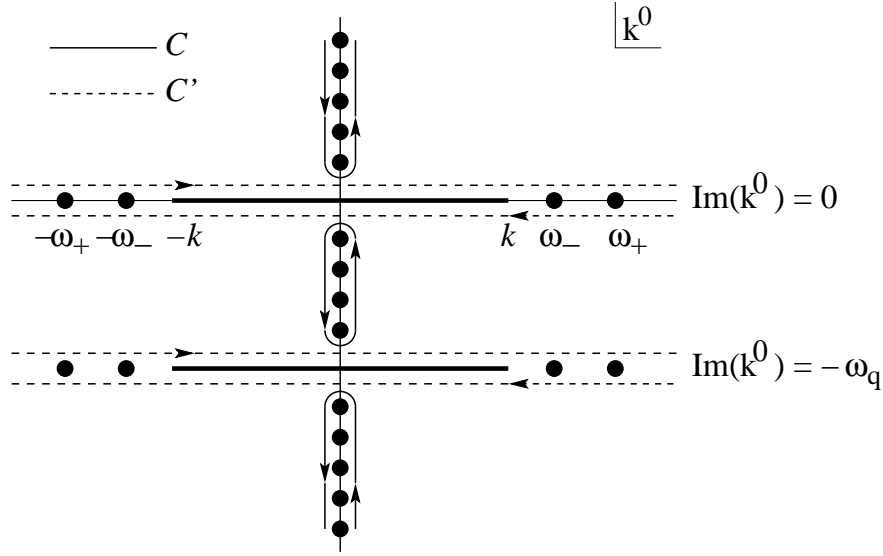}}
 \end{center}
 \caption{Contour used to do the sum over Matsubara frequencies in
 Eq.~(\ref{di}). The contour $C$ is deformed into $C'$ surrounding the
 poles and cuts. The fermionic HTL propagator $\aste S(K)$ has a branch
 cut from $-k$ to $k$ and also poles at $\om_{\pm}$ and $-\om_{\pm}$, 
 where $\om_{\pm}$ are the dispersion relations with $\om_{\pm}(k)\geq k$. 
 The HTL vertex $\aste\tw^{i}(K+Q,K)$ has the same branch cut.}
 \label{circuito}
\end{figure}

In order to do the Matsubara frequency sum we follow the steps of 
Ref.~\cite{valle}, using the contour depicted in Fig.~\ref{circuito}.
After doing the analytic continuation (\ref{ac}), we arrive at
\bea    \nn
 \Gamma^i_{\rm HTL} &&\hm =  e^{2}\di_{\kv,\om}
 \int\frac{du}{2i\pi}n_{F}(u)\gamma^{\mu}\Bigg[ 
 \\ \nonumber &&
 \aste S^{R}(u+\qo,\kv)\aste\tw^{i}(u+\qo+i0^{+},u+i0^{+};\kv)\aste 
 S^{R}(u,\kv)\frac{1}{u-(\po-\om)+i0^{+}}
 \\ \nonumber &&
 -\aste S^{R}(u+\qo,\kv)\aste\tw^{i}(u+\qo+i0^{+},u-i0^{+};\kv)\aste 
 S^{A}(u,\kv)\frac{1}{u-(\po-\om)+i0^{+}}
 \\ \nonumber &&
 +\aste S^{R}(u,\kv)\aste\tw^{i}(u+i0^{+},u-\qo-i0^{+};\kv)\aste 
 S^{A}(u-\qo,\kv)\frac{1}{u-(\po+\qo-\om)-i0^{+}}
 \\ \nonumber &&
 -\aste S^{A}(u,\kv)\aste\tw^{i}(u-i0^{+},u-\qo-i0^{+};\kv)
 \aste S^{A}(u-\qo,\kv)\frac{1}{u-(\po+\qo-\om)-i0^{+}}
 \Bigg]
 \\ \nonumber && \times
 \gamma^{\nu}P^{T}_{\mu\nu}(\vvuni)\rho_{T}^{\rm free}(\om,\vvuni)
 \\ \nonumber &&\hm
 + e^{2}\di_{\kv,\om} n_{B}(-\om)
 \gamma^{\mu}\aste S^{R}(\po+\qo-\om,\kv)
 \aste\tw^{i}(\po+\qo-\om+i0^{+},\po-\om-i0^{+};\kv)
 \\ \nonumber && \times 
 \aste S^{A}(\po-\om,\kv)
 \gamma^{\nu} P^{T}_{\mu\nu}(\vvuni)
 \rho_{T}^{\rm free}(\om,\vvuni) 
 \\ \nonumber &&\hm
 + e^{2} \di_{\kv,\om} \gamma^{\mu} \Bigg[
 \aste S^{R}(\ompm+\qo,\kv)
 \aste\tw^{i}(\ompm+\qo,\ompm;\kv)h_{\pm}(\kvuni)Z_{\pm}(\kv)
 \frac{n_{F}(\ompm)}{\po-\om-\ompm}
 \\  \nonumber && +
 h_{\pm}(\kvuni)Z_{\pm}(\kv)\aste\tw^{i}(\ompm,\ompm-\qo;\kv)
 \aste S^{A}(\ompm-\qo,\kv)
 \frac{n_{F}(\ompm-\qo)}{\po+\qo-\om-\ompm}
 \\ \nonumber &&
 + \aste S^{R}(-\ompm+\qo,\kv)
 \aste\tw^{i}(-\ompm+\qo,\ompm;\kv)h_{\mp}(\kvuni)Z_{\pm}(\kv)
 \frac{n_{F}(-\ompm)}{\po-\om+\ompm}
 \\  \nonumber &&
 + h_{\mp}(\kvuni)Z_{\pm}(\kv)\aste\tw^{i}(-\ompm,-\ompm-\qo;\kv)
 \aste S^{A}(-\ompm-\qo,\kv)
 \frac{n_{F}(-\ompm-\qo)}{\po+\qo-\om+\ompm}\Bigg] 
 \\ && \times 
 \gamma^{\nu}P^{T}_{\mu\nu}(\vvuni)\rho_{T}^{\rm free}(\om,\vvuni),
 \label{vertexz}
\eea
where $\vv=\pv-\kv$. The first group of terms come from the branch cuts, 
the second from the pole of the photon propagator
and the last group (which must be written for $\omp$ and $\omm$) come from 
the poles of the HTL propagators.
After doing the integral in $u$ we arrive at
\bea 
 \nn
 \tw^{i}_{\rm HTL} &&\hm =
 e^{2}\di_{\kv,\om} \left[n_{F}(\omega)+n_{B}(\omega-p)\right] 
 \gamma^{\mu}\aste S^{R}(\om+\qo,\kv)   
 \\ \nn && \times
 \aste\tw^{i}(\om+\qo+i0^{+},\om-i0^{+};\kv)
 \aste S^{A}(\om,\kv)\gamma^{\nu}P^{T}_{\mu\nu}(\vvuni)
 \rho_{T}^{\rm free}(p-\om,\vvuni)  
 \\ \nn  &&\hm
 + e^{2}\di_{\kv,\om} \gamma^{\mu} \Bigg[
 \aste S^{R}(\ompm+\qo,\kv) 
 \aste\tw^{i}(\ompm+\qo,\ompm;\kv) h_{\pm}(\kvuni) Z_{\pm}(\kv)
 \frac{n_{F}(\ompm)}{\om-\ompm}  
 \\ \nn  &&
 + h_{\pm}(\kvuni)Z_{\pm}(\kv) \aste\tw^{i}(\ompm,\ompm-\qo;\kv)
 \aste S^{A}(\ompm-\qo,\kv)
 \frac{n_{F}(\ompm-\qo)}{\om+\qo-\ompm}
 \\ \nn &&
 + \aste S^{R}(-\ompm+\qo,\kv) 
 \aste\tw^{i}(-\ompm+\qo,\ompm;\kv)h_{\mp}(\kvuni)Z_{\pm}(\kv)
 \frac{n_{F}(-\ompm)}{\om+\ompm}
 \\  \nn &&
 + h_{\mp}(\kvuni)Z_{\pm}(\kv)\aste\tw^{i}(-\ompm,-\ompm-\qo;\kv)
 \aste S^{A}(-\ompm-\qo,\kv)
 \frac{n_{F}(-\ompm-\qo)}{\om+\qo+\ompm}\Bigg] 
 \\ &&
 \times \gamma^{\nu} P^{T}_{\mu\nu}(\vvuni) 
 \rho_{T}^{\rm free}(p-\om,\vvuni),
 \label{res}
\eea
where we have made the change of variable $p-\om\to\om$. We are interested 
in computing the real part after
contracting with the spinors $\bar{u}_\lambda(\pvuni)\dots 
u_\lambda(\pvuni)$. First 
we must show that if we expand in
the external frequency $\qo$, the real part of the first term $1/\qo$ 
vanishes. For the last group of terms
we notice that since $\om_{\pm}(k)\geq k$ both the vertex and the 
propagators are real (apart from the Dirac
matrix structure). With the help of
\bea
 \nn &&\hm 
 \aste S^{R/A}(\ompm\pm\qo,\pv) =
 \frac{Z_{\pm}(\pv)h_{\pm}(\pvuni)}{\pm\qo}+\mbox{regular terms},  
 \\ &&\hm 
 \aste S^{R/A}(-\ompm\pm\qo,\pv) = 
 \frac{Z_{\pm}(\pv)h_{\mp}(\pvuni)}{\pm\qo}+\mbox{regular terms}, 
 \\ \nn &&\hm 
 \aste\tw^{\mu}(\pm\om_{\pm}+\qo,\pm\om_{\pm};\kv)]\big|_{\qo=0} = 
 \aste\tw^{\mu}(\pm\om_{\pm},\pm\om_{\pm}-\qo;\kv)\big|_{\qo=0},
\eea
it is easy to see that the last group of terms is regular when $\qo$ 
vanishes. Now, using the spectral density of
the transverse photon Eq.~(\ref{rsf}), we see that the integral over $\om$ 
is restricted to be below the light-cone. Therefore the propagators do not 
have poles and so there are no
pinching poles. After doing the algebra the first term in Eq.~(\ref{res}), 
contracted with $\puni^i$, can be written as
\bea \nn
 e^{2}\int_{\kv,\om} &&\hm 
 \left[n_{F}(\omega)+n_{B}(\omega-p)\right] 
 \rho_{T}^{\rm free}(p-\om,\vvuni) 
\\  \nn  \times 
 \Big[  &&\hm (A+B+C)
 \aste\Delta^{R}_{+}(\om+\qo,\kv)\aste\Delta^{A}_{+}(\om,\kv)
 \,\pvuni\cdot\kvuni (1-\kvuni\cdot\vvuni\pvuni\cdot\vvuni)    
\\ \nn  + &&\hm (A-B-C) 
 \aste\Delta^{R}_{-}(\om+\qo,\kv)\aste\Delta^{A}_{-}(\om,\kv)
 \,\pvuni\cdot\kvuni(1+\kvuni\cdot\vvuni\pvuni\cdot\vvuni)
\\ +
 &&\hm B \left\{
 \aste\Delta^{R}_{+}(\om+\qo,\kv)\aste\Delta^{A}_{-}(\om,\kv)
 +
 \aste\Delta^{R}_{-}(\om+\qo,\kv)\aste\Delta^{A}_{+}(\om,\kv)
 \right\}
 \pvuni\cdot\vvuni(\pvuni\cdot\kvuni\kvuni\cdot\vvuni-\pvuni\cdot\vvuni)
 \Big].\nn
 \\ &&\hm 
\eea
The HTL vertex has an imaginary part below the light-cone which is
singular when $\qo$ vanishes. However, the three combinations of
propagators in the previous formula are real when $\qo\to0$, and since 
we need the real part of the new diagram, we also only
need the real part of the HTL vertex. This is finite when the external
frequency goes to zero, hence the real part of the previous formula has
the same property. Thus for the real part of Eq.~(\ref{res}) we can safely
put $\qo$ to zero. Finally, since there are no pinching poles, which could
cancel some of the powers of the coupling constant, we conclude that the
whole expression is smaller than the tree level vertex contribution. After 
writing $\om = kx$, expanding in powers of $k/p$ and making the scaling 
$\kv/m_{f} = {\bf y}$, it can easily be seen that the contribution of
Eq.~(\ref{res}) behaves as $e^{3}$ times a finite integral independent of 
the coupling constant. Therefore the real part of the new diagram is, in 
the pinching-pole limit, suppressed by $e^3$ and it can safely be 
neglected.

\end{document}